\documentclass[review]{elsarticle}

\usepackage{lineno,hyperref}
\modulolinenumbers[5]

\journal{Journal of \LaTeX\ Templates}

\usepackage{times}
\usepackage{subfig}
\usepackage{soul}
\usepackage{url}
\usepackage{hyperref}
\usepackage[utf8]{inputenc}
\usepackage{amsfonts}
\usepackage{graphicx}
\usepackage{amsmath}
\usepackage{booktabs}
\usepackage{algorithm}
\usepackage{algorithmic}
\usepackage{multirow}
\usepackage{xspace}
\usepackage{footnote}% Use the postscript times font!
\usepackage{color}

\let\oldhat\hat
\renewcommand{\vec}[1]{\mathbf{#1}}
\renewcommand{\hat}[1]{\oldhat{\mathbf{#1}}}
\renewcommand{\matrix}[1]{\mathbf{#1}}

\usepackage{epstopdf}
\usepackage{threeparttable}
\usepackage{dcolumn}
\newcolumntype{d}[1]{D{.}{.}{#1}}% or D{.}{,}{#1} or D{.}{\cdot}{#1}

\newcommand{\eg}{\emph{e.g.,}\xspace}

\newcommand{\ie}{\emph{i.e.,}\xspace}

\newcommand{\etal}{\emph{et al.}\xspace}

\newcommand{\eat}[1]{}
\newcommand{\paratitle}[1]{\vspace{1ex}\noindent \textbf{#1}}

\begin{document}

%
% The "title" command has an optional parameter,
% allowing the author to define a "short title" to be used in page headers.
\title{Person-job fit estimation from candidate profile and related recruitment history with Co-Attention Neural Networks}

\author[HUST]{Ziyang Wang}
% \ead{ziyang1997@hust.edu.cn}

\author[HUST]{Wei Wei\corref{correspondingAuthor}}
\cortext[correspondingAuthor]{Corresponding author}
\ead{weiw@hust.edu.cn}

\author[HUST,JS]{Chenwei Xu}

\author[Gaoling]{Jun Xu}

\author[BIT]{Xian-Ling Mao}

\address[HUST]{School of Computer Science and Technology, Huazhong University of Science and Technology, China}
\address[Gaoling]{Gaoling School of Artificial Intelligence, Renmin University of China}
\address[BIT]{School of Computer, Beijing Institute of Technology}
\address[JS]{State Grid Jiangsu Electric Power Co., Ltd. Information \& Telecommunication Branch, China}

% \author{Chenwei Xu\fnref{a}}
% \author{Wei Wei\fnref{a}}
% \author{Ziyang Wang\fnref{a}}
% \author{Jun Xu\fnref{b}}
% \author{Xianling Mao\fnref{c}}


\begin{abstract}
Existing online recruitment platforms depend on automatic ways of conducting the person-job fit, 
whose goal is matching appropriate job seekers with job positions. 
Intuitively, the previous successful recruitment records contain important information, which should be helpful for the current person-job fit. Existing studies on person-job fit, however, mainly focus on calculating the similarity between the candidate resumes and the job postings on the basis of their contents, without taking the recruiters' experience~(\ie historical successful recruitment records) into consideration. 
In this paper, we propose a novel neural network approach for person-job fit, which estimates \textbf{p}erson-\textbf{j}ob \textbf{f}it from candidate profile and related recruitment history with \textbf{c}o-\textbf{a}ttention \textbf{n}eural \textbf{n}etworks (named PJFCANN).
Specifically, given a target resume-job post pair, PJFCANN generates local semantic representations through co-attention neural networks and global experience representations via graph neural networks. 
The final matching degree is calculated by combining these two representations. In this way, the historical successful recruitment records are introduced to enrich the features of resumes and job postings and strengthen the current matching process. Extensive experiments conducted on a large-scale recruitment dataset verify the effectiveness of PJFCANN compared with several state-of-the-art baselines.
The codes are released at: \url{https://github.com/CCIIPLab/PJFCANN}.
\end{abstract}

% \begin{keyword}
% Recruitment Analysis \sep Person-Job Fit \sep Graph Neural Network
% \end{keyword}

\maketitle

\section{Introduction}
With the rapid development of the Internet, online recruitment platforms have also expanded rapidly. 
According to a recent report from BuisnessOfApps\footnote{\url{https://www.businessofapps.com/data/linkedin-statistics/}}, by the end of the year 2019, there will be 660 million users spread over 200 countries and 20 million jobs listed on LinkedIn.
%\footnote{\url{https://www.linkedin.com}}.
The large numbers of job candidates and job postings on the Internet cost the recruiters a lot of effort to find a suitable job. 
According to SHRM's report\footnote{\url{https://www.shrm.org/hr-today/trends-and-forecasting/research-and-surveys/documents/2017-talent-acquisition-benchmarking.pdf}}, by the end of the year 2018, recruiters need to averagely spend 36 days and 4,400 dollars for filling a job position with the right talent. Therefore, it has become an essential task that automatically matches jobs with suitable candidates, called ``Person-Job Fit''.

% \begin{figure}[!t]
% 	\centering
% 	\includegraphics[width=0.8\textwidth]{figures/motivation.pdf}
% 	\caption{Example: The impacts of Discrete Context Dependencies.}
% 	\label{fig:motivation}

% \end{figure}

% The main contributions of this work are as follows.
% %To sum up, our contributions are as follows:
% \begin{itemize}
% \item We identify the problem of modeling discrete context dependencies in sequence labeling tasks.
% \item We propose a novel \emph{position}-aware self-attention to incorporate three different positional factors
% for exploring the \emph{relative} position information among tokens;
% and also develop a self-attentional context fusion within a novel neural architecture
% to provide complementary context information on the basis of Bi-LSTM for better modeling 
% the discrete context dependencies over tokens.
% %by incorporating three different positional factors to explore the \emph{relative} position information among tokens and develop a self-attentional context fusion layer to provide complementary context information on the basis of Bi-LSTM.
% %\item A new neural architecture for sequence labeling tasks is introduced to better model discrete context dependencies and extract context features of tokens.
% \item Extensive experiments on \emph{part-of-speech} (\textbf{POS}) \emph{tagging}, \emph{named entity recognition} (\textbf{NER}) and \emph{phrase chunking} tasks verify the effectiveness of our proposed model.
% \end{itemize}
Due to its high practical value, Person-job fit has received increasing research interest~\cite{xu2018measuring,lee2007fighting,paparrizos2011machine,zhang2014research,malinowski2006matching}.
Among these researches, a typical approach is to convert the task to a supervised text matching problem, which aims at calculating the matching degree based on the text content of job postings and resumes. 
More recently, with the extensive application of deep learning invariant fields, end-to-end neural networks have been proposed to understand semantic representations for calculating the matching degrees, including CNN-based model~\cite{zhu2018person}, RNN-based model~\cite{qin2018enhancing} and the model combining RNN and  CNN~\cite{bian2019domain}.
% PJFNN~\cite{zhu2018person} which proposes a CNN-based end-to-end model; APJFNN~\cite{qin2018enhancing} which proposes a RNN-based end-to-end model adding attention strategies; and a deep
% global match network~\cite{bian2019domain} which models that combined word-level RNN and sentence-level CNN.

Despite significant progress in person-to-job matching, existing methods only focus on modeling the textual semantic information of target resumes and job postings.
However, extensive experience gained from past successful recruits is also essential, further supporting current recruiting.
% However, seasoned recruiters already have extensive experience from previous successful hires, which can further support current engagements.
For example, seasoned recruiters typically evaluate hiring from two perspectives when considering a candidate's suitability for a job.
On the one hand, recruiters consider whether the context of the resume~(\eg experience and skills) fits the context of the job posting~(\eg competency requirements).
On the other hand, the experience of historically successful recruitment can be used to guide current recruitment.
The historical records can show which experiences are more important for the current job posting and which abilities requirements are more in line with the current resume.
Therefore, we propose a novel perspective to model both the text semantic matching and the experience of recruiters for person-job fit. 
%
% , as shown in the right part of Figure~\ref{motivation_example}, the obtained experience contains two part, one part is the historical employed resumes related to the current job posting, the similar work experiences in these resumes can be compared with the current resume's work experiences; Another part is the offer job postings the current candidate has already received, the requirements in these job postings can be compared with the current job posting's requirements. For this purpose, we introduce new component to model the experience of recruiters on the basis of text semantic matching component. 

Inspired by the success of graph neural network~(\ie GNN) based methods~\cite{wu2019session,wang2019neural} in the recommendation system, 
we apply GNN to incorporate the recruiters' experience into person-job fit problem. 
% We consider each recruitment as a single session and treat each session as a subgraph. 
Specifically,
by finding out all related historical successful records, we first obtain the related resumes and job postings.
Then, two graphs are constructed for each recruitment, where the obtained resumes and job postings are treated as the nodes of graphs.
% Then, two graphs are constructed for each recruitment according to the obtained resumes and job postings, where both current resume and job posting and related resumes and job postings are as nodes.
Based on the graph we build, GNN can capture the relations between the current item node and the related historical item nodes and generate node embedding vectors correspondingly. 
Finally, the latent node representation is used to generate the global embedding, which is considered as the hidden features of experiences gained from the historical recruitment records. 

Our model presents a novel perspective on modeling in the person-job fit task.
To the best of our knowledge, we are the first to model the experience of recruiters by introducing graph neural networks into the person-job fit task. 
We evaluate our proposed model on a large-scale real-world data set, and extensive experiment results have demonstrated the effectiveness of our model. 

% \textbf{Overview.} The rest of this paper is organized as follows. We briefly describe some related works in Section 2. In Section 3, we give a formal definition of person-job fit task. In Section 4, we introduce the technical details of our model. In Section 5, we compare our model performance with the latest methods on a real-world dataset. Finally, We give our conclusion in Section 6.

% \paratitle{Roadmap}. The remaining of the paper is organized as follows.
% %
% In Section \ref{sec:related}, we review the related work,
% and in Section \ref{sec:prostate} we give a formal definition of person-job fit task. We introduce the technical details of our model in Section \ref{sec:method}. Section \ref{sec:exp} presents the quantitative results on benchmark datasets, also includes an in-depth analysis and wraps up discussion over the obtained results.
% Finally, Section \ref{sec:conclusion} concludes the paper.
 
%presents some quantitative results on benchmark datasets to demonstrate the effectiveness of our proposed model. 
%It further presents an in-depth analysis through ablation study and case study and wraps up discussion over the obtained results. Section \ref{sec:related} gives an overview of the related work and Section \ref{sec:conclusion} concludes the paper.
\section{Problem Formulation}
\label{sec:prostate}
In this paper, we focus on improving the accuracy of the person-job fit task, which aims at measuring the matching degree between the job description posted by a company and the resume submitted by an employee. 

In this work, we use the experience of historical successful job applications to guide the current job application. For this purpose, we introduce the ``historical successful recruitment record". We will define the  ``historical successful recruitment record" and the task of person-job fit, respectively. 
Important notations and their corresponding definitions are shown in Table \ref{notation}.

Following~\cite{qin2018enhancing}~\cite{bian2019domain}, we use $J$ to denote a \textbf{job posting} which have $m$ ability requirements~(e.g., \textsl{C++ programming and Data Mining skills, Team Work, Communication Skill and Sincerity}). Moreover, each ability requirement $j_l$ is assumed to contain $o_l$ words.

\begin{table}[t]
\centering
\caption{Notations and Definitions.}
\label{tbl-symbol-and-definitionn}{
\begin{tabular}{c||p{220pt}} \hline
\textbf{Notation} & \textbf{Definition}   \\ \hline \hline
        $J = \{j_1, j_2, ... , j_m\}$ & A job posting which contains $m$ ability requirements. \\
        $j_l = \{j_{l,1}, j_{l,2},...,j_{l, o_{l}}\}$ & An ability requirement $j_l$ which contains $o_l$ words. \\
        $R = \{r_1, r_2, ..., r_n\}$ & A resume which have $n$ pieces of experience. \\
        $r_k = \{r_{k,1}, r_{k,2}, ... , r_{k, o_{k}}\}$ & A piece of experience $r_k$ which contains $o_k$ words. \\
        $P=\{{p_1},{p_2},...,{p_k}\}$ & $k$ job applications in the total set of job applications $P$. \\
        $Y$ & The recruitment result of this job application. \\
        ${p_i}=({J_i},{R_i},{Y_i})$ & A job application which consists of a job posting $J$, a resume $R$ and a recruitment result $Y$. \\
        $P_h$ & The set of historically successful recruitment records. \\
        \hline
\end{tabular}}
\label{notation}
\end{table}

Similarly, we use $R$ to denote a \textbf{resume} which have $n$ pieces of experience (\eg \textsl{education experiences, competition experiences, research paper publications and job experiences}). Moreover, each piece of the experience $r_k$ is assumed to contain $o_k$ words.

We use $P$ to indicate the total set of job applications along this line. Each ${p_i}$ in $P$ indicates a job application composed by a $J$-$R$ pair,  and a label $Y$ indicates the recruitment result of this job application, \ie $Y=1$ means the resume successfully passed the job description request, and the candidate entered the interview session. While $Y=0$ means the failed one. Thus, ${p_i}$ can be denoted as ${p_i}=({J_i},{R_i},{Y_i})$. Significantly, the same $J$ may exist in many different job applications, and so does $R$. However, the same $J$-$R$ pair has one and only one in $P$.

Next, we use ${P_h}$ to indicate the set of historically successful recruitment records, which is used as an experience to guide a new job application. Thus, ${P_h}$ is a subset of $P$ with each label $Y$ in ${P_h}$ is equal to 1. We define ${P_h}=\{(J,R,Y)\ |\ (J,R,Y) \in P\ and\ Y=1\}$, where ${P_h} \subset {P}$. 

Finally, we give the definition of the person-job fit task. Given a target set ${P_t}$, ${P_t} \subset {P}$ and a historical successful recruitment record set ${P_h}$, where ${P_h}\cap{P_t}=\emptyset$. 
The goal of the person-job fit task is to learn a predictive model $M$. Under the guidance of ${P_h}$, $M$ predicts a label $\oldhat{Y}$ for each $J$-$R$ pair in ${P_t}$ by calculating the matching degree between each $J$ and $R$ in ${P_t}$.
\begin{equation}
    \oldhat{M} = \arg\min \limits_{M} loss( M(\oldhat{Y}|{P_h}, {P_t}), Y)
\end{equation}
% \input{section/2_preliminaries}
% \section{Proposed Approach}
% \label{sec:method}

%\setlength{\abovecaptionskip}{-5ex}
%\setlength{\belowcaptionskip}{-5ex}
% \begin{figure}[!t]
% %\hspace{-0.4cm}
%   \centering
% 	\includegraphics[width=0.8\textwidth]{figures/figure2_new.pdf}
% 	\caption{Overview of proposed neural architecture.}
% 	\label{fig:overview}
% \end{figure}

%{\color{red}The neural architecture of our proposed model is visualized in Figure \ref{fig:overview}.}

\section{THE PROPOSED MODEL}
\label{sec:method}

\begin{figure*}[t!]
  \centering
  \vspace{-5mm}
  \includegraphics[width=\linewidth]{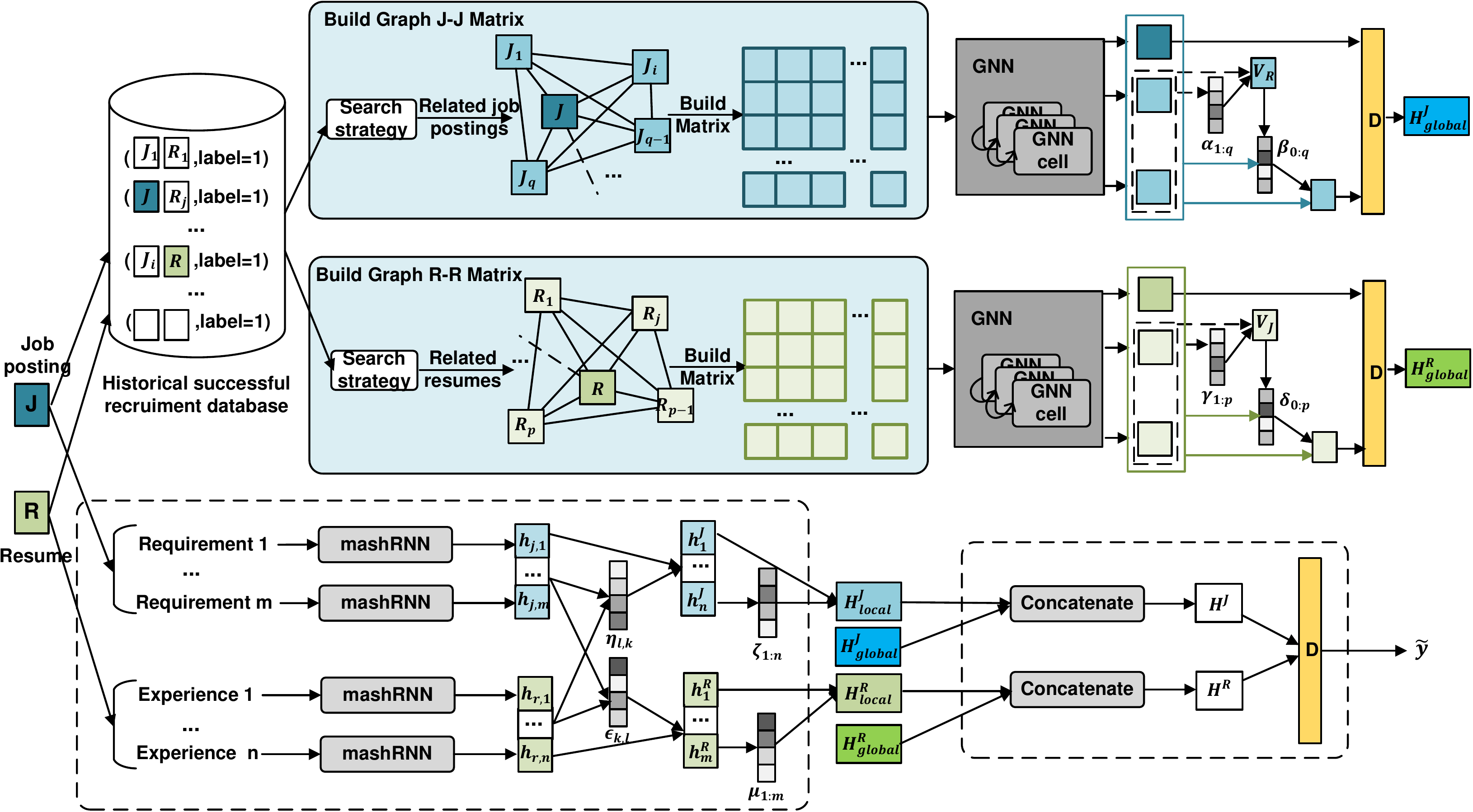}
  \caption{An illustration of the proposed PJFCANN, which consists of three parts: Self-Semantic Representation, Historical Recruitment Experience Representation and Person-Job Fit Label Prediction.}
  \label{PJFGNN}
\end{figure*}

In this section, we will illustrate the details of our model. 
As shown in Figure~\ref{PJFGNN}, PJFCANN consists of three parts, namely \emph{self-semantic representation}
, \emph{historical recruitment experience guidance} and \emph{person-job fit label prediction}.

\subsection{Self-Semantic Representation}

Self-semantic representation aims at calculating the semantic similarity between job postings and resumes. As we mentioned in the introduction, it is unreasonable to only consider job postings and resume as simple text. 

After obtaining the documents of users' requirements and experience, it is essential to use a neural network to extract the textual features from documents. A simple way is to use a recurrent neural network~(RNN). However, RNN mainly focuses on short sentences while dealing with long-form documents. 
Thus we choose mashRNN~\cite{jiang2019semantic}, which learns the document representations from multiple abstraction levels of the document structure and is capable of dealing with the long-form document. Compared with LSTM and GRU, mashRNN learns a more comprehensive semantic representation of the entire document via the information in different structure levels.
\begin{equation}
\begin{aligned}
    \vec{h}_{j,i} &= \text{mashRNN}(\matrix{E}^J_i), \\
    \vec{h}_{r,i} &= \text{mashRNN}(\matrix{E}^R_i), 
\end{aligned}
\end{equation}
% \begin{equation}
% \begin{aligned}
%     h^{j}_i &= \text{mashRNN}(e^{j_i}_{1:m}), \\
%     h^{r}_i &= \text{mashRNN}(e^{r_i}_{1:n}), 
% \end{aligned}
% \end{equation}
where $\matrix{E}^J_i$ and $\matrix{E}^R_i$ denote the matrix of word vectors for the $i$-th requirement in job posting and the $i$-th experience in resume, respectively. $\vec{h}_{j,i} \in \mathcal{R}^{d}$ and $\vec{h}_{r,i}  \in \mathcal{R}^{d}$ are the output of mashRNN and can be regarded as the semantic feature representation of the $i$-th ability requirement in job posting and the $i$-th experience in resume, respectively. $d$ is the dimension of the vectors.

After generating the feature vector, we calculate the matching degree between each requirement and each experience. 
To capture the semantic similarity between job postings and resumes, co-attention neural networks are employed to quantify the matching contributions of each candidate experience to a specific ability requirement. 

% Formally, for the $l$-th candidate experience $r_l$, its semantic feature representation $h_{r_l}$ is calculated by mashRNN. 
Here we use an attention-based relation score $\eta_{l,k}$ to quantify the matching contribution of each ability requirement $J_{k}$ to each candidate experience $R_l$, which can be calculated as follows,
\begin{equation}
\begin{aligned}
    e_{l, k} &= \vec{v}^{\rm T}_{1}\operatorname{tanh}(\matrix{W}_{1} \vec{h}_{r,l} +\matrix{U}_{1}\vec{h}_{j,k}), \\
    \vec{\eta}_{l,k} &= \frac{\exp(e_{l, k})}{\sum_{k=1}^{m}\exp(e_{l, k})}, \\
    \vec{h}^J_l &= \sum_{k=1}^{m}\vec{\eta}_{l,k} \vec{h}_{j,k},
\end{aligned}
\end{equation}
where $\matrix{W}_{1} \in \mathcal{R}^{d \times d}$ and $\matrix{U}_{1} \in \mathcal{R}^{d \times d}$, $\vec{v}_{1} \in \mathcal{R}^{d \times 1}$ are trainable parameters and  $\vec{h}^J_l \in \mathcal{R}^{d}$ is the learnt semantic feature representation of ability requirement.
Similar to relation score $\eta_{l,k}$, we can also obtain the relation score $\epsilon_{k, l}$ to quantify the matching contribution of each candidate experience $R_{l}$ to the each ability requirement $\vec{J}_{k}$ as follows,
\begin{equation}
\begin{aligned}
    e_{k, l} &= \vec{v}^{\rm T}_{2}\operatorname{tanh}(\matrix{W}_{2}\vec{h}_{j,k} + \matrix{U}_{2}\vec{h}_{r,l}), \\
    \epsilon_{k, l} &= \frac{exp(e_{k, l})}{\sum_{l=1}^{n}exp(e_{k, l})}, \\
    \vec{h}^R_k &= \sum_{l=1}^{n}\epsilon_{k, l} \vec{h}_{r,l},
\end{aligned}
\end{equation}
where $\matrix{W}_{2} \in \mathcal{R}^{d \times d}$, $\matrix{U}_{2} \in \mathcal{R}^{d \times d}$ and $\vec{v}_{2} \in \mathcal{R}^{d \times 1}$ are trainable parameters and $\vec{h}^R_k \in \mathcal{R}^{d}$ is the learnt semantic feature representation of experience.

% And for the k-th ability requirement $j_k$, its semantic feature representation $h_{j_k}$ is also calculated by mashRNN. We also use an attention-based relation score $\epsilon_{k,l}$ to quantify the matching contribution of each semantic representation $h_{j_k}$ to the l-th candidate experience $r_l$.

Then we add another attention layer to learn the importance of each representation of requirement $\vec{h}^J_l$ and each representation of experience $\vec{h}^R_k$ respectively.
Specifically, we calculate the importance $\zeta_l$ of each requirement representation $\vec{h}^J_l$ to generate the final local job posting vector $\matrix{H}^J_{local}$ as follows,

\begin{equation}
\begin{aligned}
    c^J_l &= \vec{v}^{\rm T}_{\zeta}\operatorname{tanh}(\matrix{W}_{3}\vec{h}^J_l + \vec{b}_{3}), \\
    \zeta_{l} &= \frac{exp(c^J_l)}{\sum_{l=1}^{n}exp(c^J_{l})}, \\
    \matrix{H}^J_{local} &= \sum_{l=1}^{n}\zeta_{l}\vec{h}^J_l,
\end{aligned}
\end{equation}
where $\vec{v}^{\rm T}_{\zeta} \in \mathcal{R}^{d \times 1}$, $\matrix{W}_{3} \in \mathcal{R}^{d \times d}$ and $\vec{b}_{3} \in \mathcal{R}^{d}$ are trainable parameters.
$\matrix{H}^J_{local} \in \mathcal{R}^{d}$ are the obtained local semantic vectors of job postings. And we can also calculate the importance $\mu_k$ of each candidate experience $\vec{h}^R_k$ to generate the final local resume vector $H^R_{local}$ as follows,
\begin{equation}
\begin{aligned}
    c^R_k &= \vec{v}^{\rm T}_{\mu}\operatorname{tanh}(\matrix{W}_{4}\vec{h}^R_k + \vec{b}_{4}), \\
    \mu_{k} &= \frac{exp(c^R_k)}{\sum_{k=1}^{m}exp(c^R_k)}, \\
    \matrix{H}^R_{local} &= \sum_{k=1}^{m}\mu_{k}\vec{h}^R_{k},
\end{aligned}
\end{equation}
where $\vec{v}^{\rm T}_{\mu} \in \mathcal{R}^{d \times 1}$, $\matrix{W}_{4} \in \mathcal{R}^{d \times d}$ and $\vec{b}_{4}  \in \mathcal{R}^{d}$ are trainable parameters. $\matrix{H}^R_{local} \in \mathcal{R}^{d}$ are the obtained local semantic vectors of resumes.

\subsection{Historical Recruitment Experience Representation}

% \begin{figure*}[t!]
%   \centering
%   \vspace{-5mm}
%   \includegraphics[width=0.91\linewidth]{pic/build_matrix.pdf}
%   \caption{An illustration of the process of building Job posting-Job posting Matrix and Resume-Resume Matrix}
% \vspace{-2mm}
%   \label{SEARCH}
%   \vspace{-2mm}
% \end{figure*}

\vspace{1mm}
In this part, we focus on modeling the experience of recruiters to guide the current recruitment by using historical successful recruitment records $P_h$. 
% To be specific, we use $J_c$ and $R_c$ to denote the job posting and resume of the current pair respectively. First, we search in $P_h$ to find all job postings related to $J_c$ and all resumes related to $R_c$. And then, we build \textbf{Graph J-J} based on the relation between the retrieved job postings and $J_c$, we also build \textbf{Graph R-R} based on the relation between the retrieved resumes and $R_c$. After this, we consider $J_c$ and $R_c$ as single node in their respective graphs, and learn the node embedding in each graph through the graph neural networks (GNN) and model the experience of recruiters based on attention strategies. 
We illustrate the details of \emph{building graph matrix}, \emph{learning the node embedding} and \emph{modeling the experience of recruiters}. 

% % \vspace{1mm}
% \paratitle{$\bullet$ Building Graph Matrix}. We build two adjacent matrix for each recruitment, called \textbf{Graph J-J} and \textbf{Graph R-R} respectively. Graph J-J is the graph of the relations between the historical job postings and the current job posting $J_c$. Similarly, Graph R-R is the graph of the relations between the historical hired Resumes and the current Resume $R_c$. 

\subsubsection{Building Graph Matrix} 
For each recruitment, we build two undirected graphs, named \textbf{Graph J-J} and \textbf{Graph R-R}, respectively. As from Figure \ref{PJFGNN}, Graph J-J $G_J = {(V_J, E_J)}$ contains the historical job postings and current job posting $J_c$ as nodes and the set of edges $E_J$ representing their relations. Similarly, Graph R-R $G_R = {(V_R, E_R)}$ contains the historical hired Resumes and current Resume $R_c$ as nodes and the set of edges $E_R$ representing their relations. We described the details of constructing the Graph J-J as follows and the construction process of Graph R-R is similar.

For the current recruitment pair $(J_{c}, R_{c})$, the first step to build Graph J-J is to find all job postings related to $J_c$. The search strategy is finding a set of
job posting $J_{r}$ where $J_i \in J_r$ meet the condition $(J_{i} ,R_{c}, 1) \in P_h$. After searching, each $J_i \in J_r$ is marked as the related job postings of current job posting $J_c$. 

The second step is calculating the edge weight to construct an adjacency matrix for each graph. 
{
First, we consider $J_c$ and the related job postings $[J_1, J_2, ..., J_q]$ found by search strategy as different single nodes, we first obtain the representations based on BiLSTM with attention mechanism~\cite{zhou2016attention} and then use the \emph{cosine similarity function} to calculate the similarity between two nodes to construct the adjacency matrix $A^J$,
}
\begin{equation}
    \matrix{A}^J[1][2] = cosine \left( BiLSTM(J_1), BiLSTM(J_2) ) \right),
\end{equation}
where BiLSTM denotes the BiLSTM with attention mechanism from \cite{zhou2016attention} and cosine denotes the cosine similarity function.
Note that we compute edge weight for every two nodes in the graph, which means the constructed graph is a complete graph.

\begin{figure*}[t!]
  \centering
  \vspace{-5mm}
  \includegraphics[width=\linewidth]{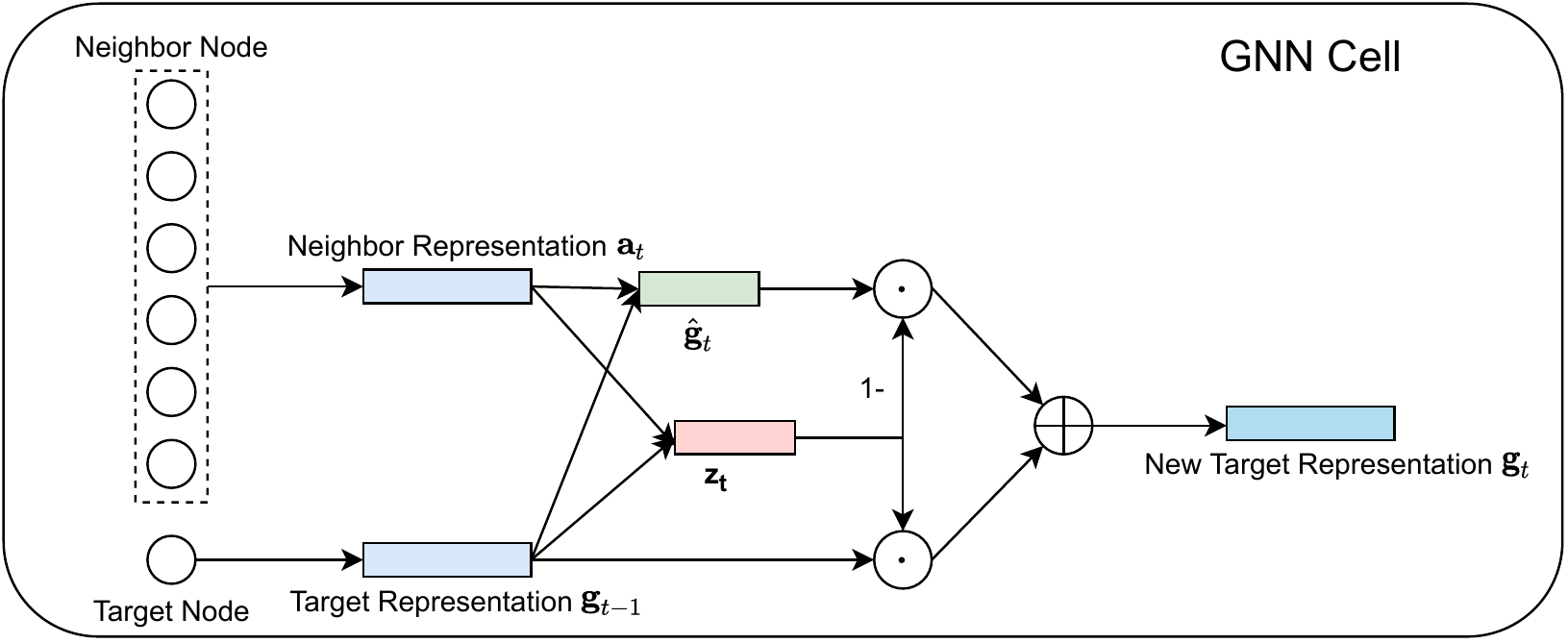}
  \caption{An illustration of the GNN cell. In the GNN cell, we first aggregate the neighbor nodes to generate the neighbor representation $\vec{a}_t$, then the new target representation $\vec{g}_t$ is obtained based on the neighbor representation $\vec{a}_t$ and original target representation $\vec{g}_{t-1}$.}
  \label{GNNCell}
\end{figure*}

\subsubsection{Learning Node embedding on Graph Matrix} 
After building the graph matrix, we get two matrices $A^J$ and $A^R$, which consider each job posting in Graph J-J and each resume in Graph R-R as a single node.
As shown in Figure \ref{GNNCell}, the node representations are learnt by the update functions of GNN cell as follows:
\begin{equation}
\begin{aligned}
    \vec{a}^{t}_i &= \matrix{A}_i([\vec{g}^{t-1}_0, \ldots, \vec{g}^{t-1}_n]^\mathrm{T} \matrix{H}^t + \vec{b}^t), \\
    \vec{z}^t_i &= \sigma (\matrix{W}^t_z \vec{a}^t_t + \matrix{M}^t_z \vec{g}^{t-1}_{i}), \\
    \vec{r}^t_i &= \sigma (\matrix{W}^t_r \vec{a}^t_t + \matrix{M}^t_r \vec{g}^{t-1}_{i}), \\
    \tilde{\vec{g}^t_t} &= \operatorname{tanh} (\matrix{W}^t_h \vec{a}^t_t + \matrix{M}^t_h(\vec{r}^t_i \odot \vec{g}^{t-1}_{i})), \\
    \vec{g}^t_i &= (1-\vec{z}^t_i) \odot \vec{g}^{t-1}_{i} + \vec{z}^t_i \odot \tilde{\vec{g}^t_i},
\end{aligned}
\end{equation}
where  $[ g^t_0, \ldots, g^t_n ] \in \mathcal{R}^{d \times n}$ is the list of node vectors in current label graph matrix at layer $t$, $A_i \in \mathcal{R}^{1 \times n}$ is $i$-th row of the matrix corresponding to node $i$. 
$H^t$, $W^t_z$, $M^t_z$, $W^t_r$, $M^t_r$, $W^t_h$, $M^t_h \in \mathcal{R}^{d \times d}$ and $b^t \in \mathcal{R}^{d}$ are the parameters to learn. $z^t \in \mathcal{R}^{d}$ and $r^t \in \mathcal{R}^{d}$ are the reset and update gates respectively. $\odot$ is element-wise multiplication. $\sigma (\cdot)$ is the logistic function. 
Note that we construct a trainable job posting embedding matrix $\matrix{H}^{GJ} \in \mathcal{R}^{|J| \times d}$ and resume embedding matrix $\matrix{H}^{GR} \in \mathcal{R}^{|R| \times d}$ here. 
And the initial representations of the nodes~(\ie $\vec{g}^{0}$) are obtained by looking up the embedding matrices $\matrix{H}^{GJ}$ and $\matrix{H}^{GR}$.

After feeding two matrices into the respective gated graph neural network, we obtain the representations of all nodes in Graph J-J and Graph R-R, respectively,
which are denoted as $\vec{g}^J$ and $\vec{g}^R$, respectively. Among these, $\vec{g}^J_0$ and $\vec{g}^R_0$ indicate the representation of input job posting and resume (\ie $J$ and $R$ in the Figure \ref{PJFGNN}), respectively.

\subsubsection{Modeling the Experience of recruiters} 
After getting the final node vectors in Graph J-J and Graph R-R, we further extract more high-level representations for modeling the experience of recruiters. The recruiter knows precisely which ability in resumes he successfully passed is a strong match for this job posting and which requirement in job postings has a high matching degree with this resume. We define the two experiences above as relation J-R and R-J, respectively.

Relation J-R focuses on enhancing the modeling of candidate abilities. First, we apply the soft-attention mechanism to map the job postings (\ie historical successful recruitment records for the current resume) embeddings to Graph R-R vector space.

\begin{equation}
\begin{aligned}
\alpha_i &= {\vec{v}_j}^\mathrm{ T } \sigma(\matrix{W}_j \vec{g}^J_i + \vec{c}), \\
\matrix{V}_R &= \sum _ { i = 1 } ^ { q } \alpha _i \vec{g}^J_i, \\
\end{aligned}
\end{equation}
where $\vec{g}^J_i$ denotes each node vector in the Graph J-J, parameters $\vec{v}_j \in \mathcal{R}^{d \times 1}$ and $\matrix{W}_j \in \mathcal{R}^{d \times d}$ denotes the trainable parameters. $\matrix{V}_R \in \mathcal{R}^{d}$ denotes the representation of job postings in Graph R-R vector space. It estimates the outstanding skills that recruiters have valued in past successful hires. We use another attention mechanism to estimate the matching degree between each job posting and $\vec{V}_R$.

\begin{equation}
\begin{aligned}
f^R_{t} &= \vec{v}_{\beta}^{\rm T} \operatorname{tanh}(\matrix{W}_{\beta}\vec{V}_R+\matrix{U}_{\beta}\vec{g}^J_t), \\ 
\beta_{t} &= \frac{exp(f^R_{t})}{\sum_{t=0}^{q}exp(f^R_{t})}, \\
\vec{e}^J &= \sum_{t=0}^{q}\beta_{t}\vec{g}^J_t, \\ 
\end{aligned}
\end{equation}
where $\vec{v}_{\beta} \in \mathcal{R}^{d \times 1}$, $\matrix{W}_{\beta} \in \mathcal{R}^{d \times d}$ and $\matrix{U}_{\beta} \in \mathcal{R}^{d \times d}$ are the parameters to be learned during the training processing. The attention score $\beta$ can be seen as the matching degree between requirements in job postings and ability in current resume. $\vec{e}^J$ denotes the experience from relation J-R. 

Finally, we concatenate the job posting representation $\vec{g}^J_0$ from Graph J-J with the experience representation $e^J$ from relation J-R and consider the output as the global vector of the resume.

\begin{equation}
\begin{aligned}
\matrix{H}^J_{global} &= \operatorname{tanh}(\matrix{W}_J[\vec{g}^J_0; \vec{e}^J] + \vec{b}_J), \\
\end{aligned}
\end{equation}
where $\matrix{W}_J  \in \mathcal{R}^{d \times 2d}$ and $\vec{b}_J \in \mathcal{R}^{d}$ are the parameters to learn.

% As we present the process of learning the global feature vector of relation J-R $H^J_{global}$, the global feature vector of relation R-J $H^R_{global}$, which focuses on modeling the recruiter's personal preferences, is learned in the same way as $H^J_{global}$.
Relation R-J focus on modeling the recruiter's personal preferences. 
First we apply the soft-attention mechanism to map presentation of the resumes (\ie historical successful recruitment records for the current job posting) embeddings to Graph J-J vector space.

\begin{equation}
\begin{aligned}
\gamma_i &= {\vec{q}_r}^\mathrm{ T } \sigma(\matrix{W}_r \vec{g}^R_i + \vec{c}), \\
\vec{V}_J &= \sum _ { i = 1 } ^ { p } \alpha _i \vec{g}^R_i, \\
\end{aligned}
\end{equation}
where $\vec{g}^R_i$ denotes each node vector in the Graph R-R, parameters $\vec{q}_r \in \mathcal{R}^{d \times 1}$ and $\matrix{W}_r \in \mathcal{R}^{d \times d}$ denotes the trainable parameters.
$\vec{V}_J$ denotes the representation of resumes in Graph J-J vector space, and estimates the outstanding skills that recruiters have valued in past successful hires. We use another attention mechanism to estimate the matching degree between each resume and $\vec{V}_J$.

\begin{equation}
\begin{aligned}
    f^J_{t} &= \vec{v}_{\delta}^{\rm T}tanh(\matrix{W}_{\delta}\vec{V}_J+\matrix{U}_{\delta}\vec{g}^R_t), \\ 
    \delta_{t} &= \frac{exp(f^J_{t})}{\sum_{t=0}^{q}exp(f^J_{t})}, \\
    e^R &= \sum_{t=0}^{p}\delta_{t}g^R_t, \\ 
\end{aligned}
\end{equation}
where $\vec{v}_{\delta} \in \mathcal{R}^{d \times 1}$, $\matrix{W}_{\delta} \in \mathcal{R}^{d \times d}$ and $\matrix{U}_{\delta} \in \mathcal{R}^{d \times d}$ are the parameters to be learned during the training processing. The attention score $\delta$ can be seen as the matching degree between requirements in job postings and ability in current resume.
$\vec{e}^R$ denotes the experience from relation R-J. Finally, we concatenate the resume representation from Graph R-R $\vec{g}^R_0$ with the experience representation from relation R-J $\vec{e}^R$, and consider the output as the global information vector of the current resume.

\begin{equation}
\vec{H}^R_{global} = \operatorname{tanh}(\matrix{W}_R[\vec{g}^R_0; \vec{e}^R] + \vec{b}_R),
\end{equation}
where $\matrix{W}_R \in \mathcal{R}^{d \times 2d}$ and $\vec{b}_R \in \mathcal{R}^{d}$ are the parameters to learn.

\subsection{Person-Job Fit Label Prediction}
After we get the local semantic vectors $\vec{H}^J_{local}$, $\vec{H}^R_{local}$ and the global information vectors $\vec{H}^J_{global}$, $\vec{H}^R_{global}$, we first concatenate the local vectors and the global vectors. Then, a comparison mechanism is applied based on a fully connected network to measure the matching degree between the job posting and the resume. Finally, we send the output $\vec{D}$ of the fully connected network into a logistic function to get the predicted label $\Tilde{y}$.

\begin{equation}
\begin{aligned}
\vec{H}^J &= [\vec{H}^J_{local}; \vec{H}^J_{global}], \\
\vec{H}^R &= [\vec{H}^R_{local}; \vec{H}^R_{global}], \\
\vec{D} &= \operatorname{tanh}(\matrix{W}_d[\vec{H}^J; \vec{H}^R; \vec{H}^J-\vec{H}^R] + \vec{b}_d), \\
\oldhat{Y} &= \sigma (\matrix{W}_y\vec{D} + \vec{b}_y),
\end{aligned}
\end{equation}
where $\matrix{W}_d  \in \mathcal{R}^{d_2 \times 6d}$, $\vec{b}_d  \in \mathcal{R}^{d_2}$, $\matrix{W}_y  \in \mathcal{R}^{1 \times d_2}$, $\vec{b}_y  \in \mathcal{R}^{1}$ are the parameters to learn and $\oldhat{Y} \in [0, 1]$ is the prediction result. $\sigma( \cdot )$ is the logistic sigmoid function.
To optimize our model, we adopt the binary cross-entropy loss over the entire training data as the total loss. 

% We test different distribution and different proportion of dimension, detailed experimental result are shown in subsection 5.7.

% To learn the model parameters, we adopt the Adam optimizer\cite{kingma2014adam}, and we also adopt the dropout strategy to avoid overfitting. More detailed training settings was shown in subsection 5.1. 

% \subsection{Discussion}
% As the key components in the model, \textsl{the size of $P_h$} and  \textsl{the similarity function} we used in constructing matrix affect the model result. Therefore, we tested the effect of the size of $P_h$, in detail, which can be found in subsection 5.4. We also tested different similarity functions, detailed experimental results are shown in subsection 5.5.

\section{Experiments}
\label{sec:exp}

In this section, we first introduce the data set and detailed experimental settings. Then, we report the performances of our model and baseline methods, followed by the analysis of the results.

\begin{table}[t]
\begin{center}
\caption{Statistics of the dataset}
\begin{tabular}{ l | r }
\hline\hline	
  Statistics & Values \\ \hline\hline
  job postings number &  21661 \\
  resumes number & 2437  \\
  \# of successful applications & 12112 \\
  \# of failed applications & 38354 \\   
  Average words per job posting & 359.44  \\
  Average words per resume & 53.51 \\
  \hline\hline
\end{tabular}
\label{statistics}
\end{center}
\end{table}

\begin{figure}
\begin{center}
  \includegraphics[width=2.3in]{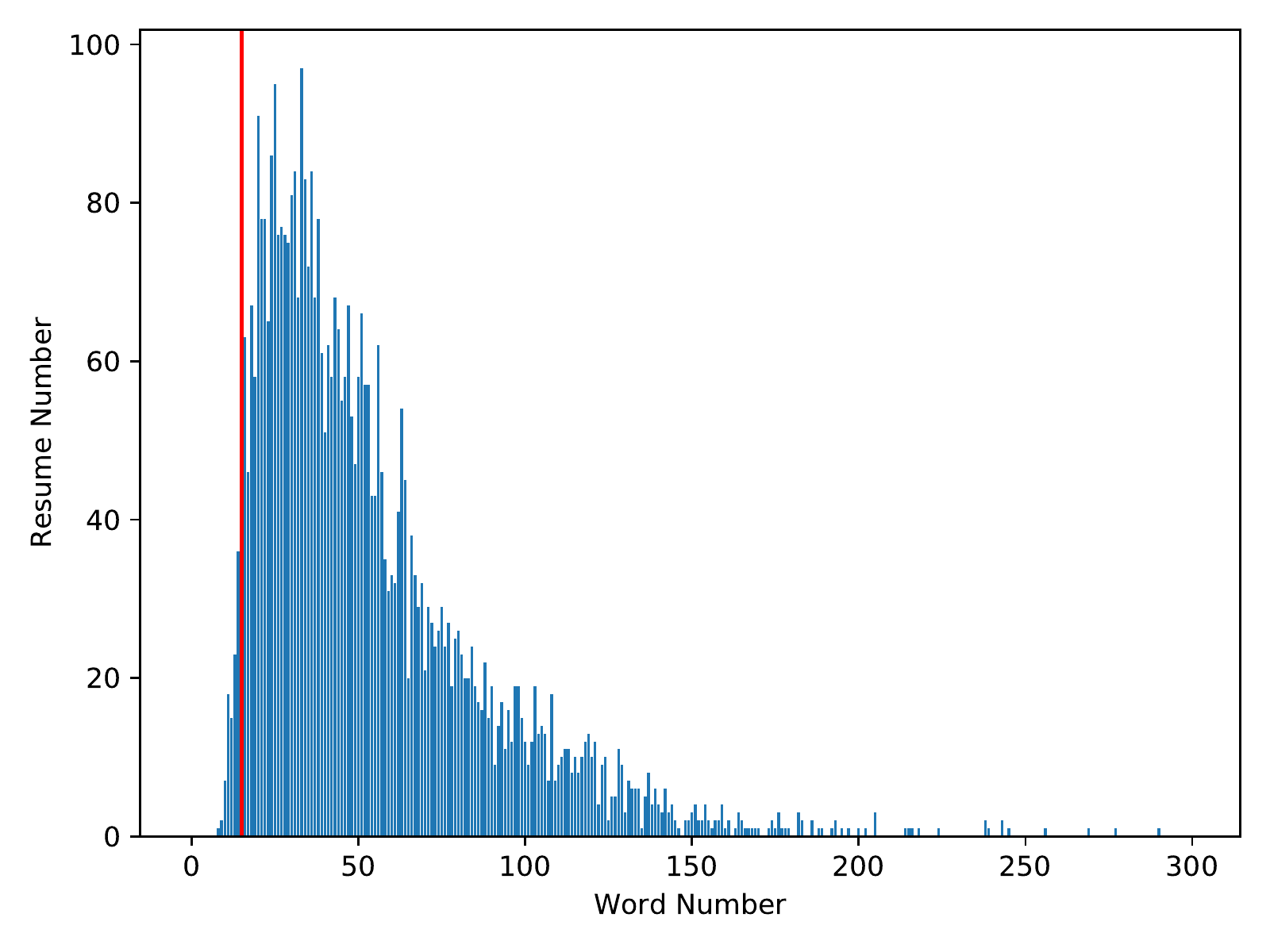}
  \includegraphics[width=2.3in]{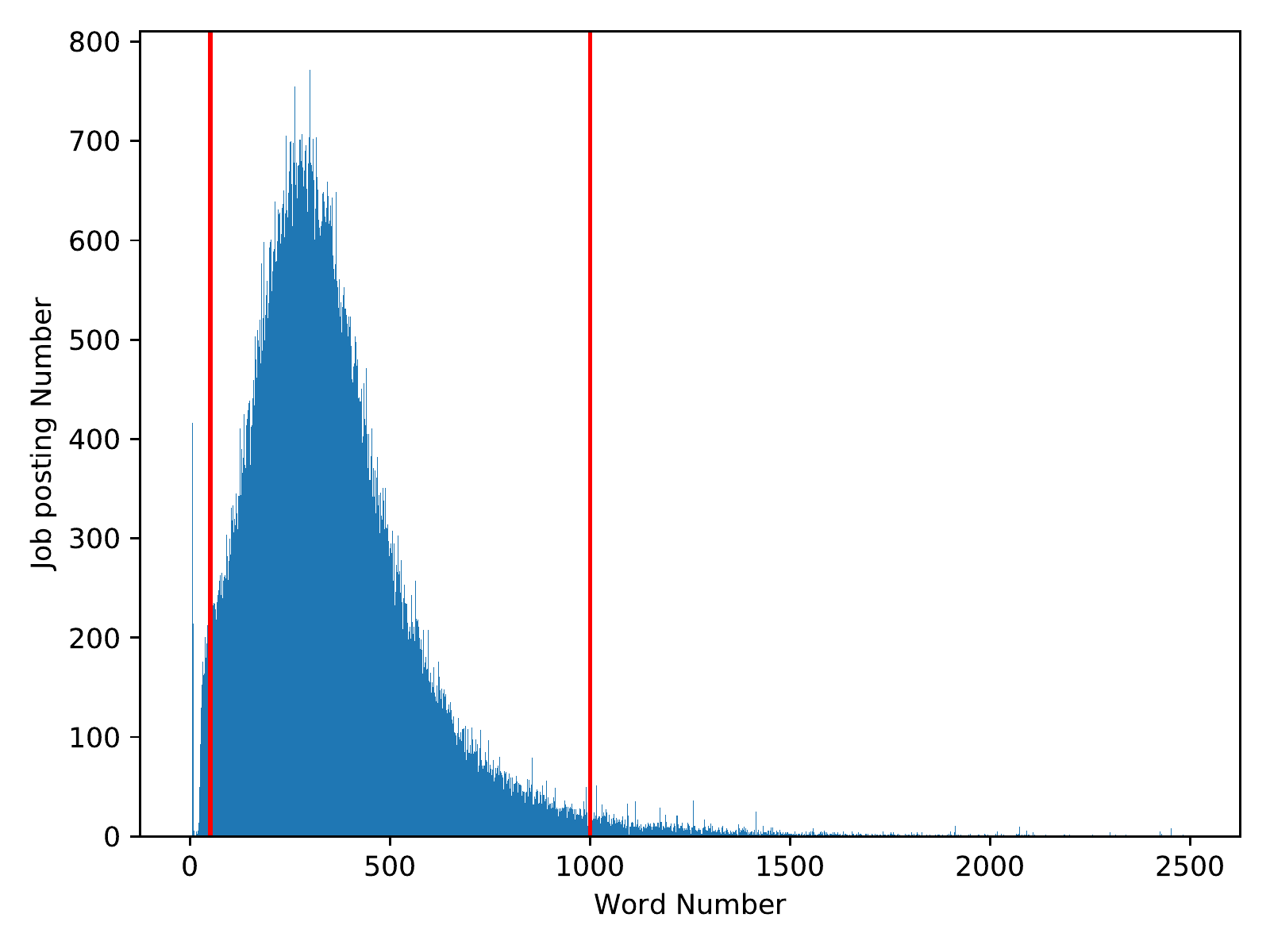}
  \caption{The words distribution of resume and job posting.}
  \label{word_distribution}
  \vspace{-3mm}
\end{center}
\end{figure}

\subsection{Experimental Settings}

We introduce the details in our experiment setting, including the data set, word embedding, settings of PJFCANN, and settings of the training stage.

\paratitle{$\bullet$ Dataset.}
We evaluate our model on a large real-world data set provided by an online recruitment company. To protect the privacy of users, all private information is anonymously processed. 

In this dataset, for a particular pair of a job posting and a resume, three labels are marking the relation between the job posting and the resume, namely \textsl{"browsed"}, \textsl{"delivered"} and \textsl{"satisfied"}. The label "browsed" means whether the user has been browsed this job posting online, the label "delivered" means whether the user has delivered his resume to this job posting online, the label "satisfied" indicates whether the user matches this job after interviews. Each label's value is chosen in $\{0, 1\}$ where 1 means yes and 0 means no. We mark the "satisfied" pairs as the successful applications and mark those pairs which are marked "delivered" but not marked "satisfied" as failed applications. 

The data set consists of 4465 user resumes, 269534 job postings, total pair number is 6971621. There are 31696 pairs marked "satisfied", 150930 pairs marked "delivered", 310521 pairs marked "browsed". According to the word distribution shown in Figure~\ref{word_distribution}, we consider those resumes whose word number is less than 15 as incomplete resumes and those job postings whose word number is less than 50 as incomplete job postings. To ensure the quality of experiments, those incomplete resumes and job postings are removed. We also set the maximum number of words in job postings as 1000, removing the excess parts. Correspondingly, those resumes and job postings without successful applications are also removed. Statistics of the pruned data are summarized in Table~\ref{statistics}.

According to the recruitment acceptance rates, there are many more failed applications than successful ones, which leads to a typical imbalanced situation. Therefore, we randomly select some failed applications to keep up with the same number of successful applications.

\begin{figure}
\begin{center}
\includegraphics[width=70mm]{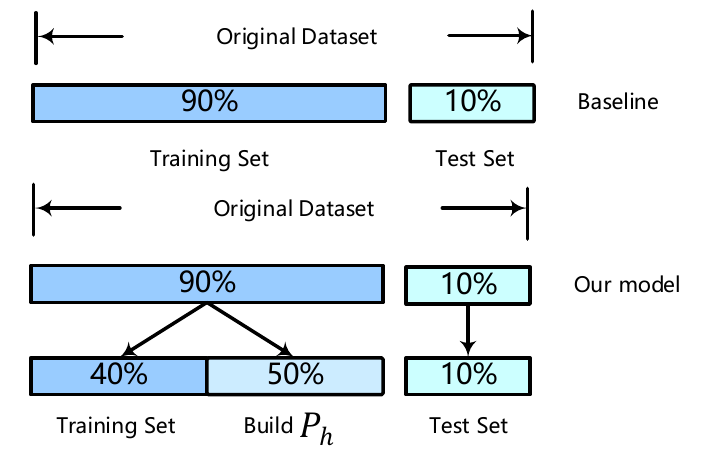}
  \caption{An illustration of the split of dataset}
  \label{dataset-split}
  \vspace{-3mm}
\end{center}
\end{figure}

Because the model needs to use external data to build the graph, for fairness, we divide the data set as follows (Figure~\ref{dataset-split}). For baseline methods, we first divide the data set into 10 pieces. In each piece, the number of successful and failed applications is equal. And we randomly select 9 pieces of the training set, the left 1 piece in the testing set. For PJFCANN, to be fair, we construct the Historical Recruitment Results $P_h$ only from the training set. We randomly select 5 pieces from the 9 pieces training set and consider the successful applications in the 5 pieces as $P_h$. The left 4 pieces are used for training, the 1 pieces testing set is still used for testing. In the training set, 10\% is used for validation, and 90\% is used for training. The reason for dividing the piece of $P_h$ and training is that we found this ratio is optimal. The experiment about the effect of $P_h$'s size is in subsection 5.4.

\paratitle{$\bullet$ Experimental parameter setting.}
We used the GloVe\footnote{https://nlp.stanford.edu/projects/glove/} to pre-train the word embedding, and used the pre-trained word embedding results to initialize the embedding layer weights. We set the dimension of word embedding as 200.

Following~\cite{jiang2019semantic}~\cite{wu2019session}, we set the dimension of hidden state in mashRNN as 1024 and the dimension of hidden state output by mashRNN $d$ is set to 200. The dimension of hidden state after fully connected network $d_2$ is set to 512. And the number of GNN layers is set to 1.

We optimize the model by using Adam~\cite{kingma2014adam} algorithm, and we set batch size as 16 for training and add the dropout layer with the probability of 0.5 to prevent overfitting. All parameters are initialized using a Gaussian distribution with a mean of 0 and a standard deviation of 0.1. The initial learning rate is set to 0.1 and will decay by 0.1 after every 2 epochs. The L2 penalty is set to $10^{-5}$. The model is trained on a server with 6-cores CPU@3.20GHz, 32GB RAM, and a GeForce GTX 1080 GPU.

We adopt the four most commonly used evaluation metrics, which are \textsl{Accuracy}, \textsl{Precision}, \textsl{Recall} and \textsl{F1 score} respectively.

\subsection{Baseline Methods}
We select some state-of-the-art supervised models as baseline methods, and we compare our model PJFCANN with these baseline methods to verify the effect of our model. 

\paratitle{$\bullet$ Classic Supervise Learning Methods.} Following \cite{qin2018enhancing}, we implement two classic supervise learning methods, \textsl{Random Forests} (RF) and \textsl{Gradient Boosting Decision Tree} (GBDT). we construct the representation by using the \textsl{Bag-of-words vectors} as the input features. 

\paratitle{$\bullet$ DSSM}\cite{huang2013learning} utilizes the deep neural network (DNN) to learn dense feature representation on job posting and resume respectively, and calculates the semantic similarity of the feature representation pair. 

\paratitle{$\bullet$ RE2}\cite{yang2019simple} is a general-purpose text-matching framework. The two text sequences are symmetrically processed before the prediction layer, and all parameters except those in the prediction layer are shared between the two sequences. We consider the resume text and the job posting text as two sequences respectively.

\paratitle{$\bullet$ PJFNN}\cite{zhu2018person} proposes a bipartite Convolutional Neural Networks to effectively learn the joint representation of resumes and job postings.

\paratitle{$\bullet$ BERT}\cite{devlin2018bert} is a language representation model which can be fine-tuned on the semantic textual similarity task. We consider the resume text and the job posting text as two sequences respectively and we fine-tuned the model based on $BERT_{BASE}$.

\paratitle{$\bullet$ BPJFNN}\cite{qin2018enhancing} is a simplified version of APJFNN. It considers resumes and job postings as two single sequences and applies Bi-LSTMs to learn the semantic representation of each word in the two sequences.

\paratitle{$\bullet$ APJFNN}\cite{qin2018enhancing} considers each experience in resume and each requirement in job postings as a sequence, and it applies four hierarchical ability-aware attention strategies based on BPJFNN to learn a word-level semantic representation for both job requirements and resumes.

\paratitle{$\bullet$ SCLPJF}\cite{bian2019domain} consists of \textsl{Single Domain part} and \textsl{Domain Adaptation part}. In the Single Domain part, it first utilizes RNN to encode the sequence, and then model global semantic interactions between sequence representations of a resume and job postings. In the Domain Adaptation part, it utilizes the SCL algorithm to learn transferable match patterns and components from a domain with sufficient labeled data to a domain without sufficient labeled data. Because our dataset did not have a domain label and can not support the Domain Adaptation part, we only compare our model with SCLPJF's Single Domain part.

\paratitle{$\bullet$ MV-CoN}\cite{bian2020learning} consist of \textsl{text-based matching model} and \textsl{relation-based matching model}. The text-based matching model is implemented by BERT and Transformer architecture. The relation-based matching model builds a job-resume relation graph and uses GNN to aggregate each node's representation. A co-teaching mechanism is used to reduce the influence of noise in training data.

\paratitle{$\bullet$ PJFCANN (w/o GNN)} is a variant of PJFCANN, which removes the $H_{global}$ part and only uses the $H_{local}$ part. Without the $H_{global}$ part, PJFCANN (w/o GNN) can not obtain the graph information~(\ie historical success recruitment records).

\subsection{Overall Results}
\begin{table}[t]
\centering
% \resizebox{\columnwidth}{!}{%
\caption{Performance of PJFCANN and baselines.}
\begin{tabular}{ l | c | c | c | c }
\hline \hline
  \textbf{Methods} & \textbf{Accuracy} & \textbf{Precision} & \textbf{Recall} & \textbf{F1}\\ \hline\hline
  RF   &	 0.6646 &	0.6783 &     0.6261 &	   0.6512 \\
  GBDT &	 0.6807 &	0.6923 &     0.6507 &	   0.6708 \\ \hline
  DSSM &	 0.6723 &	0.6830 &     0.6431 &	   0.6624 \\
  RE2  &     0.7115 &   0.7074 &     0.7215 &      0.7144 \\
  PJFNN   &  0.6938 &   0.6927 &     0.6969 &      0.6948 \\
  BERT & 0.7123& 0.7084 &     0.7215 &      0.7149 \\
  BPJFNN  &  0.7107 &   0.7051 &     0.7246 &      0.7147 \\
  APJFNN  &  0.7415 &   0.7364 &	 0.7523 &	   0.7443 \\ 
  SCLPJF~(Single Domain)  &  0.7146 &   0.7123 &    0.7200 &     0.7161 \\ 
  MV-CoN & 0.7496 & 0.7733 & 0.7063 & 0.7383\\ \hline
  \textbf{PJFCANN~(w/o GNN)} & 0.7353 & 0.7361 & 0.7338 & 0.7349 \\
  \textbf{PJFCANN} &  \textbf{0.8076} & \textbf{0.7994} & \textbf{0.8215} & \textbf{0.8103}  \\
  \hline\hline
\end{tabular}
% }
\label{baseline-result}

\end{table}

The performance is shown in Table~\ref{baseline-result}, clearly that our model outperformed all the baselines with a significant margin. 

The performance of classic supervised models~(\ie RF and GBDT) are not satisfactory, which is because they use the Bag-of-Words vectors to characterize the semantic features, whose representation ability is somehow limited compared with pre-trained word vectors. Besides, the results also show that the deep learning methods can capture the semantic features better than classic supervised models.

Then we can observe that DSSM does not perform well over four metrics because it fails to capture
the sequential properties in the textual information. RE2 and PJFNN obtain better performance than DSSM. However, they use convolutional neural networks to get the semantic features from the long-document sentences, which is still insufficient to capture sequential information in the sentences. 

Compared our model PJFCANN with BERT, PJFNN, BPJFNN, SCLPJF, and MV-CoN, it shows that our model outperforms baseline methods over four metrics. This is because these methods only focus on the textual information of current job and resume, while not fully use of the historical successful recruitment records. Although MV-CoN creates a job-resume graph based on the category labels at the bottom layer and keywords, it still suffers from two limits. Firstly, the job-resume graph based on the category labels and keywords is sparsity, which can only provide limited information. Secondly, the category labels may not exist in many real scenes, and the reliability of keywords needs to be judged due to the complexity of jobs and resumes. Unlike MV-CoN, our method uses the historical successful recruitment database to construct the job-resume graph, which is more dense and reliable. 

% The results showed the model based on deep learning is better than the classic supervise model, which indicated that it is better for fine-tuning the pre-trained word vectors than using the Bag-of-Words vectors to characterize the semantic features.

% Compared our model PJFCANN with APJFNN and SCLPJF, the better results indicated that 

Compared with all baselines above, our model had two main advantages. In the first point, our model not only learn the local representation from the text semantic information but also learn the global representation from the historical successful recruitment records. And the second point, our model had the ability to reference the experience of the recruiter, by introducing GNN, our model used the historical experience to represent the current resume and job posting, the attention mechanism help to model the recruiter's preferences and the candidate's abilities. The results indicates that our novel perspective improves the performance of the Person-Job Fit task.
Furthermore, by comparing PJFCANN and PJFCANN~(w/o GNN), we can observe that PJFCANN outperforms PJFCANN w/o GNN significantly, where the average improvements of PJFCANN are 10.16\% over PJFCANN w/o GNN in terms of four metrics. The result demonstrates the importance of historical success recruitment records $P_h$ and the superiority of our GNN module.

\begin{figure}
  \includegraphics[width=\linewidth]{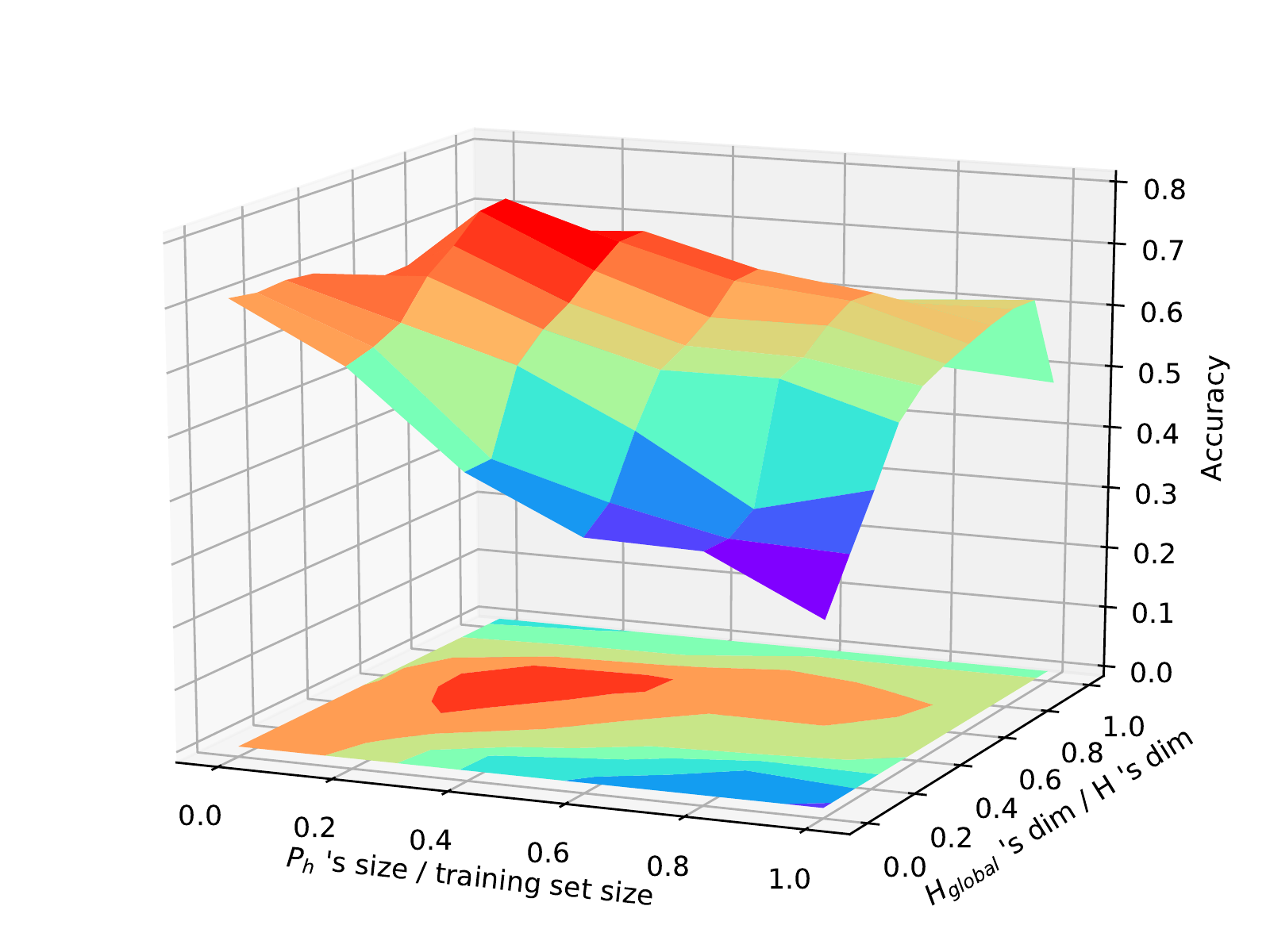}
  \caption{The effect of $P_h$'s size and $H_{global}$'s dimension to the model}
  \label{p_h-effect}
\end{figure}
\subsection{Effect of $P_h$ 's Size}
As we mention before, we divide the dataset into 10 pieces, randomly selected 9 pieces for training set, the left 1 pieces for testing set. And we randomly selected 5 pieces from the 9 pieces training set and consider the successful applications in the 5 pieces as $P_h$, the left 4 pieces is used for training. It is obviously that the size of $P_h$ and the size of training set affect the result predicted by PJFCANN.

% \begin{figure}
%   \includegraphics[width=\linewidth]{pic/permutation.pdf}
%   \caption{An illustration of the Permutation mode. (a) is the Full Permutation mode and (b) is the Random Permutation mode.}
%   \label{permutation-test-pic}
%   \vspace{-3mm}
% \end{figure}

Therefore, we test the effect of $P_h$ 's size on the model. The results are shown in Figure~\ref{p_h-effect}, where the X-axis denotes the ratio of the dimension of $H_{global}$ to the dimension of concatenation vector $H$ (\textsl{H = concat($H_{global}$; $H_{local}$})), the Y-axis denotes the ratio of $P_h$'s size to training set's size, the Z-axis denotes the accuracy tested on current model. Both the X-axis and the Y-axis's range are from 0 to 1 (including 0 and 1). The larger value of the X-axis is, the greater proportion of the vector generated by GNN in the final vector representation means that the historical experience has a greater impact on current recruitment. The larger value of the Y-axis is, the greater proportion of the historical successful recruitment records $P_h$'s size in the total training set's size means that there is more data to build the graph and less data to train text semantics.

We analyze the result from four perspectives, as shown in Figure~\ref{p_h-effect}, when x and y are both small, which means the recruiter is more focused on match text semantics of the current recruitment, the accuracy is around 68\%. When x and y are both big, the recruiter pays most attention to getting guidance from the historically successful recruitment records, the accuracy is around 63\%. This phenomenon indicates that there are limitations to modeling person-job fit tasks only from text semantics or historical recruiter's experience.

However, when x is extensive and y is small, the recruiter shows more attention to match the text semantics of the current recruitment without enough semantic training data. The accuracy is very low and only around 35\%. When x is small, and y is extensive, the recruiter pays the most attention to get guidance from the historical successful recruitment records without enough records data, the accuracy is also very low and only around 42\%. 

The highest accuracy happens when x is about 0.2 and y is about 0.6, which shows that the best results could be obtained when the semantic characteristics of the text are considered together with the historical experience.

% \subsection{Effect of Permutation Mode}
% As we mentioned in section 4.2, after searching out all related items from $P_h$, we generated all permutations of these items for computing the similarity. Therefore, we tested the effect of different permutation modes.

% As shown in Figure~\ref{permutation-test-pic}, we compared two different permutation modes, namely \textsl{full permutation mode} and \textsl{random permutation mode}. We generated all the possible permutations in full permutation mode, and randomly chose a permutation in random permutation mode.

% The results are shown in Table~\ref{permutation-test}, the result based on full permutation mode is better than random permutation mode, however, the time cost in full permutation mode is also more than random permutation mode. This phenomenon is because the sequence number generated by full permutation mode is much more than the sequence number generated by random permutation mode. 

% \begin{table}[t]
% \resizebox{\columnwidth}{!}{%
% \begin{tabular}{ l | c | c | c | c | c }
% \hline \hline
%   \textbf{Permutation} & \textbf{Accuracy} & \textbf{Precision} & \textbf{Recall} & \textbf{F1} &     \textbf{Epoch Time}\\ \hline\hline 
%   Full       & 0.8046   & 0.8010    & 0.8107   & 0.8058   & 97 mins/epoch\\
%   Random     & 0.7907   & 0.7846    & 0.8015   & 0.7929   & 64 mins/epoch\\
%   \hline\hline
% \end{tabular}}
% \caption{Effect of the permutation function.}
% \label{permutation-test}
% \vspace{-1mm}
% \end{table}

\subsection{Effect of the Encoder for Similarity Function}
As we mention in section 4.2, to get each value in the final matrix, we need a similarity function to calculate the similarity between two resumes or job postings when building a graph. Therefore, we test the effect of the different encoder of similarity functions on the model. The similarity function can be divided into two categories: \textsl{unsupervised} and \textsl{supervised}.

\noindent \textbf{$\bullet$ Unsupervised.}
We evaluate 4 unsupervised methods, namely \textsl{mean}, \textsl{tf-idf}, \textsl{smooth inverse frequency(SIF)}\cite{arora2016simple} and \textsl{Word Mover's Distance (WMD)}\cite{kusner2015word}. First, we use a word vector to represent each word in the sentence, then we obtain the corresponding sentence embedding according to the different weights of the methods. Finally, we calculate the difference between the embeddings due to the similarity.

\noindent \textbf{$\bullet$ Supervised.}
The supervised methods differ from unsupervised methods and require pretraining. For fairness, we train the supervised model based on the training set we split in subsection 5.1. 

For resume, we generate the resume pairs from the applications with the same job posting, and we label the successful application resume pair 1 indicates the high semantic similarity and label the failed application resume pair 0. we did the same label step with job posting pairs. The ratio of the training set to the valid set is 4:1.

\begin{table}[t]
\small
{
\caption{Effect of the similarity function.}
\label{similarity-func}
\begin{tabular}{ l | c | c | c | c}
\hline \hline
  \textbf{Similarity function} & \textbf{Accuracy} & \textbf{Precision} & \textbf{Recall} & \textbf{F1}\\ \hline\hline 
  PJFCANN (mean+word2vec)    &  0.6315 &	0.6359  &     0.6154 &	   0.6255 \\
  PJFCANN (tf-idf+word2vec)  &  0.6539 &	0.6793  &	  0.6323 &     0.6549 \\
  PJFCANN (WMD)              &  0.6461 &	0.6513 &	  0.6292 &	   0.6401 \\ 
  PJFCANN (SIF+word2vec)     &  0.7207 &	 0.7151 &	  0.7338 &	   0.7243 \\ \hline
  PJFCANN (CNN)              &  0.7615 &   0.7537 &     0.7769 &     0.7651 \\
  PJFCANN (BiLSTM)           &  0.7684 &   0.7722 &     0.7615 &     0.7668 \\
  \textbf{PJFCANN (BiLSTM+attention)} &  \textbf{0.8031} & \textbf{0.7949} & \textbf{0.8169} & \textbf{0.8057}  \\
  \hline
  \hline
\end{tabular}}
\end{table}

We evaluate 3 supervised methods which are all end-to-end models based on \textsl{CNN}, \textsl{BiLSTM} and \textsl{BiLSTM+attention} respectively. We use two encoders to generate two hidden representations from the two input sequences, respectively, then treat softmax's output as the value of similarity.

The testing results are shown in Table~\ref{similarity-func}. The supervised methods are better than the unsupervised methods. This phenomenon may indicate that word vectors are not enough to characterize the semantic features of the recruitment textural data. The representation after neural network's fine-tune can capture more accurate semantic information. The method based on \textsl{BiLSTM + attention} works best, which may indicate the attention strategies can capture the critical information in more detail and improve the performance with a better estimation of matching results.

\begin{table}[t]
\centering
% \resizebox{\columnwidth}{!}{%
\caption{The comparison of inference latency.}
\begin{tabular}{ l | c }
\hline 
  \textbf{Methods} & \textbf{Latency (s)}           \\ 
  \hline \hline
  BPJFNN                        &   1.624           \\
  APJFNN                        &   2.872           \\ 
  MV-CoN                        &   3.928           \\ 
  \hline
  \textbf{PJFCANN~(w/o GNN)}    &   1.826           \\
  \textbf{PJFCANN}              &   2.622           \\
  \hline
\end{tabular}
% }
\label{efficiency-result}

\end{table}

\subsection{Inference Efficiency Comparison}
The inference efficiency is essential for job-resume problem in real-world scenes. Thus we conduct experiments here to evaluate the inference efficiency of our method. Specifically, we compare the inference latency~(\ie the time cost for the inference of test data) between our method PJFCANN with selected representative methods. The evaluation is conducted on a server with 6-cores CPU@3.20GHz,  32GB RAM, a GeForce GTX1080 GPU, and batch size of 32.

The results are shown in Table \ref{efficiency-result}, it can be observed that the inference efficiency of our method is competitive by comparing with other representative baselines. In this work, we use gate graph neural network to model the historical success records, and the number of layers is set to 1. The inference latency of PJFCANN is higher than PJFCANN (w/o GNN), which demonstrates that GNN is more time-complex and needs to cost more time in inference.
However, the inference latency of PJFCANN is still lower than APJFNN and MV-CoN, which demonstrates the superiority of our model.

\begin{figure}[!t]
 \subfloat[]{
  \begin{minipage}[b]{\linewidth}
  \centering
  \includegraphics[width=0.9\linewidth]{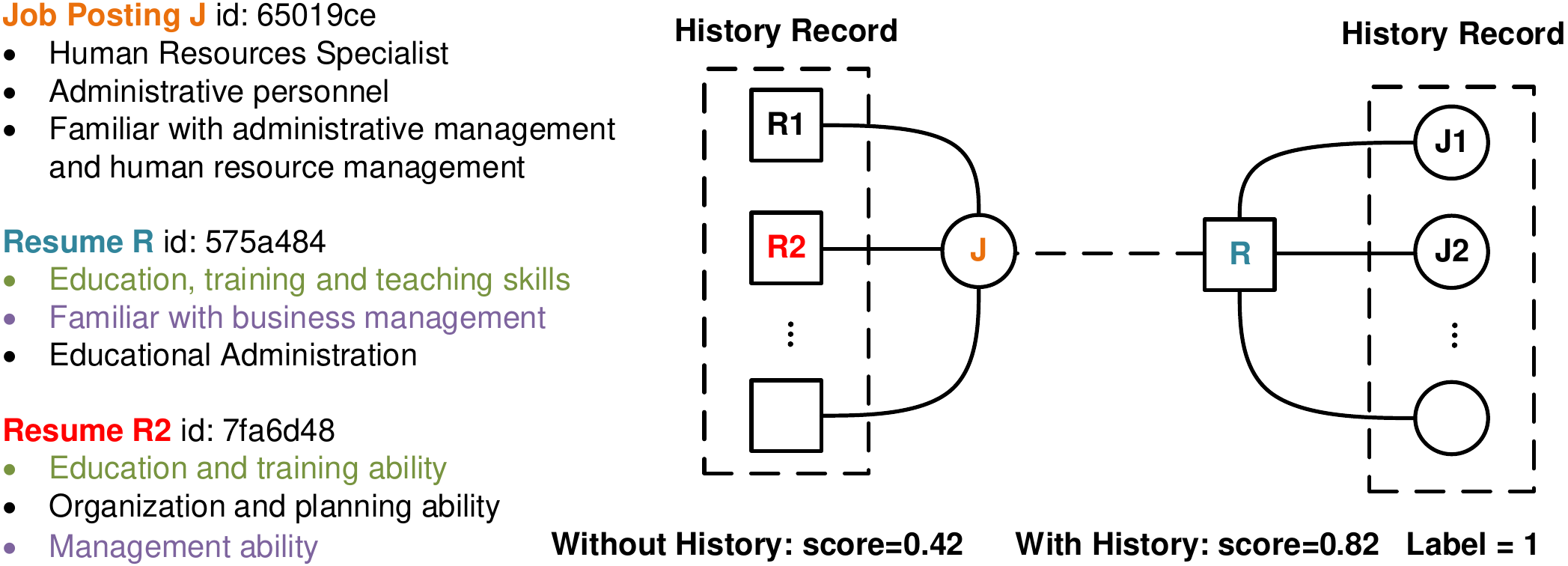}
  %\caption{fig1}
  \end{minipage}
  \label{case1}
 }

 \subfloat[]{
  \begin{minipage}[b]{\linewidth}
  \centering
  \includegraphics[width=0.9\linewidth]{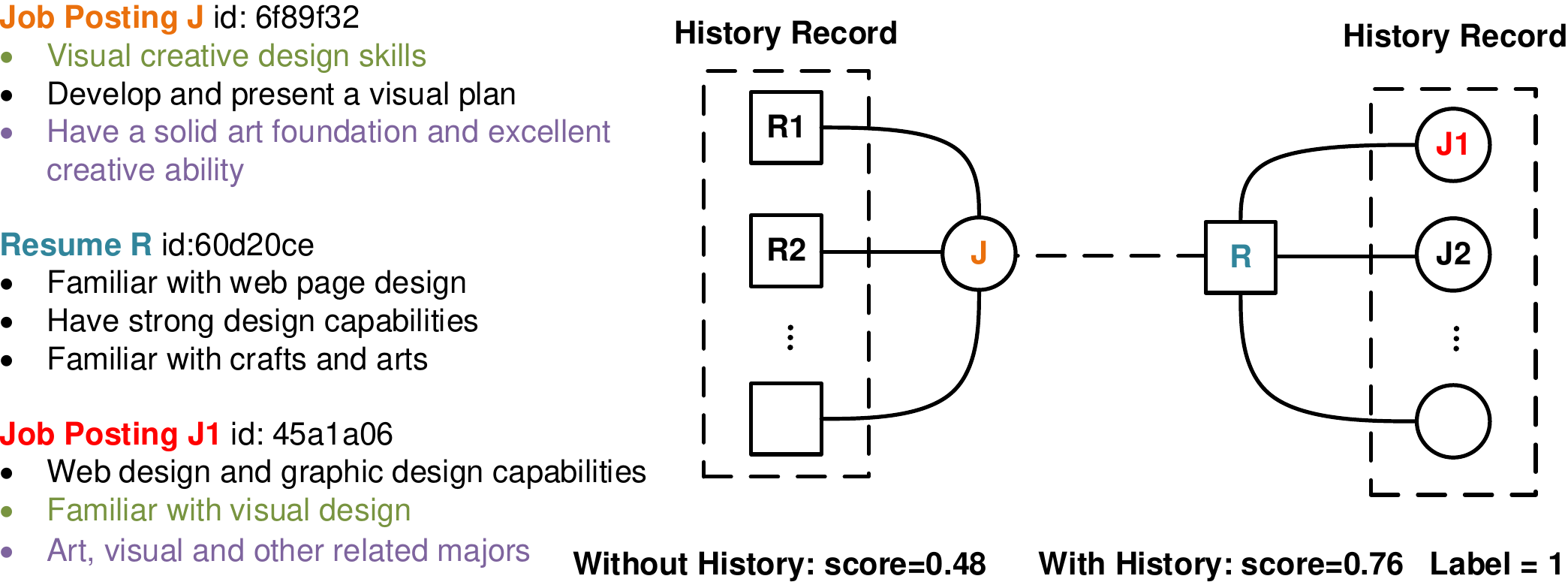}
  %\caption{fig2}
  \end{minipage}
  \label{case2}
 }
 \caption{An illustrative example for showing the effect of historical records.}
% \vspace{-0.5cm}
\label{casestudy}
\end{figure}

\subsection{Case Study}
To further show the importance of historical success recruitment records, in Figure \ref{casestudy} we present some case studies for understanding the working mechanism from a
qualitative perspective.

Form case \ref{case1}, we can observe that it is hard to directly compute the matching degree between job posting J and resume R. The job posting J focus on ``human resources specialist" and ``administrative management'' ability. However, it is difficult to directly capture the correlation between these requirements and the content on the candidate resume R. By incorporating the historical success recruitment records of job posting J and resume R, we can easily find the correlation between resume R2 and candidate resume R~(\eg Education, training and management ability). Thus the model can capture the match signals by incorporating the information of resume R2 into job posting J via GNNs. Similar observation can also be observed from the case \ref{case2}. The cases show that using simple text semantic matching is inadequate, and the historical success recruitment records are significant for job-resume matching.

\section{Related Work}
\label{sec:related}

Our work involves research in two direction: namely \textsl{Recruitment Research}, \textsl{Deep learning in Text Mining}.

\subsection{Recruitment Research}
The research on recruitment has never been stopped. With the big open-source data of online recruitment, more and more research task on recruitment analysis has emerged. \cite{wang2013time} \cite{zhu2016recruitment} research the recruitment market trend over time and when to make appropriate job recommendations to a user. \cite{cheng2013jobminer} \cite{lin2017collaborative} research the company profiling to build an in-depth understanding of the company’s fundamental characteristics. \cite{harris2017finding} \cite{javed2017large} analyze the differences in job requirements between technical recruitment and non-technical recruitment. \cite{xu2018measuring} propose a data-driven approach for modeling the popularity of job skills based on the analysis of large-scale recruitment data.

The person-job fit~\cite{sekiguchi2004person} which focuses on improving the matching degree between employers' skills and job requirements, is one striking task of recruitment analysis. Early research treats person-job fit as a recommendation problem. Malinowski~\etal~\cite{malinowski2006matching} argue that a match between a candidate and a job needs to be bilateral and apply two distinct recommendation systems to improve the matching degree. Following this idea, Lee~\etal~\cite{lee2007fighting} introduces a comprehensive job recommender system for employers to find their interesting jobs. \cite{paparrizos2011machine} first formulate this recommendation problem as a supervised machine learning problem and use naive Bayes hybrid classifier to model it. \cite{diaby2013toward,hong2013dynamic,lu2013recommender,zhang2014research} add different collaborative filtering algorithm into the recommendation system for job seeking and use it on the recruiting website. \cite{zhang2016glmix} apply the Generalized Linear Mixed Model into the job recommendation system.

Recent research begins to analyze the person-job fit task from another angle. \cite{zhu2018person} treat the task as a classification problem and propose an end-to-end model for matching text semantic between resumes and job postings based on Convolutional Neural Network(CNN). Following this idea, \cite{qin2018enhancing} keep applying BiLSTMs and Attention mechanism instead of CNN and obtain a better result. Based on this model, \cite{bian2019domain} add the Structural Correspondence Learning(SCL) algorithm to solve the problem of unbalanced data in different job categories.
MV-coN~\cite{bian2020learning} integrates both text- and relation-based matching models into a unified approach for the job-resume matching, where the relation graph is built based on category and keywords information. However, relying solely on category and keyword information to build graphs may not be reliable enough and will introduce much noise.

Unlike previous works, our proposed PJFCANN leverages the experience of historical success recruitment records to guide the current recruitment, which is necessary and reliable information to enrich the features of resumes and job postings.

\subsection{Text Mining with Neural Network}
Generally, the study of person-job fit based on textual information can be grouped into the tasks of text mining, which is highly related to Natural Language Processing (NLP) technologies, such as text classification \cite{kim2014convolutional} \cite{yang1997comparative}, text similarity \cite{gomaa2013survey} \cite{kim2014convolutional} \cite{severyn2015learning} and reading comprehension \cite{berant2014modeling} \cite{hermann2015teaching} Recently, due to the advanced performance and flexibility of deep learning, more and more researchers try to leverage
deep learning to solve the text mining problems. Compared with
traditional methods that largely depend on the effective human-designed representations and input features (e.g., word n-gram \cite{wang2012baselines}, parse trees \cite{cherry2008discriminative} and lexical features \cite{melville2009sentiment}), the deep learning-based approaches can learn effective models for large-scale textural data without labor-intensive feature engineering. Among various deep learning models, Convolutional Neural Network (CNN) \cite{lecun1998gradient} and Recurrent Neural Network (RNN) \cite{elman1990finding} are two representative and widely-used architectures, which can provide effective ways for NLP problems from different perspectives. Specifically, CNN efficiently extracts local semantics and hierarchical relationships in textural data. For instance, as one of the representative works in this field, Kalchbrenner et al. \cite{kalchbrenner2014convolutional} proposed a Dynamic Convolutional Neural Network (DCNN) for modeling sentences, which obtained remarkable performance in several text classification tasks. Furthermore, Kim et al. have shown the power of CNN on a wide range of NLP tasks, even only using a single convolutional layer \cite{kim2014convolutional}. From then on, CNN-based approaches have attracted much more attention on many NLP tasks. For example, in \cite{he2015multi}, He et al. used CNN to extract semantic features from multiple levels of granularity for measuring the sentence's similarity. Dong et al. introduced a multi-column CNN for addressing the Question Answering problem \cite{dong2015question}.

\section{Conclusions}
\label{sec:conclusion}

In this paper, we propose a novel \textbf{P}erson-\textbf{J}ob-\textbf{F}it with \textbf{Co}-\textbf{A}ttention \textbf{N}eural \textbf{N}etwork (PJFCANN) model in which the matching degree between the job postings and resumes are predicted based on not only their context but also the related successfully matched job postings and resumes in history, which reflects the recruiters' experiences in a large extent. 
%The key idea is to model the recruiter's experience while calculating text similarity. 
Specifically, given a target resume-job post pair, PJFCANN first generates the local semantic representations based on a Recurrent Neural Network (RNN). At the same time, PJFCANN generates the global experience representations for the pair based on a Graph Neural Network (GNN). Therefore, the final matching degree is calculated based on the concatenation of these two representations. In this way, the recruiters' experiences are naturally utilized in the matching process. To test the performances of the proposed PJFCANN, extensive experiments were conducted on a large-scale recruitment data set from a commercial online recruitment company. The results verified the effectiveness of PJFCANN compared with several state-of-the-art baselines.

For future work, more prior knowledge information~(\eg knowledge graph) can be incorporated to enhance the ability of text feature learning.
And a promising direction is to design more robust ways of graph construction, which can break our dependence on historical data and incorporate more semantic information.
Moreover, it is significant to improve the efficiency of the person-job fit (i.e., to implement
an online detection system).

\section{Acknowledgments}
This work was supported in part by the National Natural Science Foundation of China under Grant No.61602197 and Grant No.61772076.

%
% The next two lines define the bibliography style to be used, and
% the bibliography file.
\bibliographystyle{ACM-Reference-Format}
\bibliography{sample-base}

% \appendix

\end{document}